%% file: Automatica_draft.tex
\documentclass[twocolumn]{autart}    

\usepackage{graphicx}          

\usepackage{epsfig} 
\usepackage{mathptmx} 
\usepackage{times} 
\usepackage{amsmath} 
\usepackage{amssymb}  
\usepackage[latin1]{inputenc}
\usepackage{amsfonts}
\usepackage{makeidx}
\usepackage{graphicx}
\usepackage{epstopdf}
\usepackage{algorithm}
\usepackage{tikz}
\usetikzlibrary{shapes.geometric,arrows,patterns,fadings,positioning}

\usepackage{framed}
\usepackage[shortlabels]{enumitem}
\usepackage{scalefnt}
\usetikzlibrary{decorations.pathmorphing}

\usepackage{graphics}
\usepackage{amsmath}
\usepackage{amsfonts}
\usepackage[version=3]{mhchem}

\graphicspath{{./Images/}}

\begin{document}

\begin{frontmatter}

\title{Multiple Window Moving Horizon Estimation\thanksref{footnoteinfo}} 

\thanks[footnoteinfo]{This paper was not presented at any IFAC 
meeting. Corresponding author A.~A.~Al-Matouq. Tel. +001-720-934-3591. 
Fax +001-303-869-5009.}

\author[Ali]{Ali Al-Matouq}\ead{aalmatou@mines.edu},    
\author[Ali]{Tyrone Vincent}\ead{tvincent@mines.edu},               

\address[Ali]{Department of Electrical Engineering and Computer Science, Colorado School of Mines 1600 Illinois St., Golden, CO 80401, USA}  


\begin{abstract}                          
Long horizon lengths in Moving Horizon Estimation are desirable to reach the performance limits of the full information estimator.  However, the conventional MHE technique suffers from a number of deficiencies in this respect.  First, the problem complexity scales at least linearly with the horizon length selected, which restrains from selecting long horizons if computational limitations are present.  Second, there is no monitoring of constraint activity/inactivity which results in conducting redundant constrained minimizations even when no constraints are active.  In this study we develop a Multiple-Window Moving Horizon Estimation strategy (MW-MHE) that exploits constraint inactivity to reduce the problem size in long horizon estimation problems.  The arrival cost is approximated using the unconstrained full information estimator arrival cost to guarantee stability of the technique.  A new horizon length selection criteria is developed based on maximum sensitivity between remote states in time.  The development will be in terms of general causal descriptor systems, which includes the standard state space representation as a special case.  The potential of the new estimation algorithm will be demonstrated with an example showing a significant reduction in both computation time and numerical errors compared to conventional MHE. 
\end{abstract}

\end{frontmatter}

\begin{section}{Introduction}
Inequality constraints in estimation problems can arise from known boundaries in the dynamics of the system emerging from physical insight and can be viewed as additional a-priori information.  The added value of inequality constraints in state estimation is well known and demonstrated in many fields, see for example the studies given in \cite{Robertson2002} and \cite{Haseltine2005}.  Inequality constraints may also arise in convex filtering problems as in $\ell_{1}$ trend filtering and total variation de-noising \cite{chu2012moving} or when other densities with finite support describe the system and/or measurement noise \cite{Robertson2002}, \cite{Goodwin2005},\cite{aravkin2013optimization}.  

Unfortunately inequality constraints in the estimation problem generally prevents the use of recursive solutions for finding the estimates \cite{Rao2000}, \cite{Rawlings2009}.  The moving horizon estimate (MHE), on the other hand, is found by limiting the estimation problem to a window of measurements and system dynamic updates that slides with time while partially accounting for past measurements through an extra penalty cost term, referred to as an arrival cost \cite{Rao2000}.  The horizon length is selected based on many factors, including computational limitations, system observability and model accuracy.  Higher estimation accuracy, in terms of mean square error, may be achieved by using long horizon lengths or alternatively, finding more accurate arrival cost approximations, provided that the model uncertainties are well accounted for \cite{Rawlings2009}.  Efforts to improve arrival cost approximations can be found in \cite{Rao2000} and in \cite{chu2012moving} for linear state space systems and in \cite{Rao2000},  \cite{ungarala2009computing}, \cite{qu2009computation}, \cite{Zavala20101662} and \cite{Lopez-Negrete2011} for non-linear state space systems.

The conventional sliding window technique in Moving Horizon Estimation, however, can become computationally inefficient.  For example, at certain times the solution of the inequality constrained state estimation problem may be identical to the solution of the unconstrained problem; i.e. dropping the inequality constraints from the sliding window minimization problem for these states has no effect on the value of the estimates.  For linear systems, these unconstrained state estimates can be determined using simple recursive solutions and hence there is no added value in including these states in the minimization problem and they create an unnecessary computational burden.  Moreover, numerical errors associated with window minimizations increase with the size of the sliding window used.

Recently in \cite{chu2012moving} an approximation hypothesis was used to derive a simple arrival cost update for general staged QP problems with sufficiently large horizon lengths by assuming that the active and inactive state constraints of the last state in the moving horizon window remain respectively active or inactive indefinitely after exiting the window.  Consequently, equality constraints corresponding to the active inequality constraints were included in the arrival cost update.  However, no stability analysis was provided using this method, nor means for selecting the sufficiently large horizons.

This technique seems very attractive but can cause problems when the horizon is not chosen large enough to satisfy the active constraint hypothesis.  If short horizons are used, for example, then estimator divergence may result if the presumably indefinite active constraint is not really active after smoothing the state (i.e. after more measurements are collected).  This overweighting of past data may result in neglecting new data and potentially can cause estimator divergence if the coupling between the states in time is strong \cite{Rawlings2009}.    On the other hand, dropping inequality constraints from the minimization problem once they pass outside the window has no destabilizing effect, as the estimator will possess the guaranteed convergence and stability properties of the unconstrained estimator.  

In view of the above, a numerical algorithm that accounts for active constraints over large horizons without compromising stability or efficiency is developed.  This is achieved by using an arrival cost approximation that guarantees stability by exploiting regions of constraint inactivity to automatically reformulate the objective function into a reduced form.  A complete convergence and stability analysis for our selection of the arrival cost for descriptor MHE is given in the appendix using analogies with the presentation given in \cite{Rao2000} and \cite{Rawlings2009} for state space systems.  

Figure \ref{fig1} shows an explanation of our new proposed strategy that enables efficient long horizon estimations.  A short sliding window objective function is used to scan for states with active constraints.  The inequalities associated with states that never became constrained inside the sliding window are dropped from future minimizations assuming they will remain unconstrained.  On the other hand, inequalities associated with states that were constrained inside the sliding window objective remain in subsequent minimizations and form fixed windows that are augmented to the sliding window.  These fixed windows, (within the intervals $[a_{1},b_{1}]$ and $[a_{2},b_{2}]$ as shown in Fig. \ref{fig1}) remain in the estimation problem until their influence on the current state is negligible.  In effect, the objective function is adaptively modified according to the activity of constraints while exploiting constraint inactivity to reduce problem complexity.  The algorithm was implemented using the semi-definite programming solver \cite{toh1999sdpt3} with the CVX parser in Matlab \cite{gb08} and is available on-line.

\input{fig1}

In conventional MHE, the horizon length is often selected based on computational limitations because of the linear growth of problem complexity with horizon length.  If the system to be estimated, however, operates mostly inside the region defined by the inequality constraints, and intermittently operates near the constraints, than the new MW-MHE can be used to exploit regions of constraint inactivity to perform long horizon estimation efficiently.  This also promotes selecting horizon lengths based on the sensitivity between remote states in time rather than based on implementation restrictions.  Hence, a new tuning method for selecting the horizon length based on a user specified minimum magnitude of acceptable coupling between distant states in time is also developed in this study.

The development will be in terms of general causal descriptor systems, which includes the standard state space representation as a special case.  Our motivation for descriptor systems is estimation problems that involve differential algebraic models ubiquitous in simulation environments \cite{biegler2012control}, and problems that involve singularly perturbed systems \cite{kumar1999}.  Descriptor systems have been also used in unknown input estimation in \cite{Darouach1995} which avoids improvising a random walk model on the input signal.  Moreover, other staged QP filtering and estimation problems can benefit from the descriptor system framework, like $\ell_{1}$ trend filtering and total variation de-noising \cite{chu2012moving}.  Descriptor moving horizon estimation was first considered in \cite{Boulkroune2010}.  

The paper is organized as follows. Section 2 presents the constrained full information estimation problem for descriptor systems followed by the required assumptions.  The Moving Horizon approximation is then presented in Section 3 following the theme given in \cite{rao2001constrained}, where the relationship between full information and moving horizon estimation was analysed using dynamic programming.  Section 4, which is the main contribution of this study, will present the multiple window moving horizon estimation algorithm and a new tuning parameter based on the coupling between remote states in time.  Finally, Section 5 will demonstrate the potential of the new MW-MHE algorithm with an example.  The appendix sections details the  proofs of theorems used in this study. 

The following notation is used in this study: $\mathbb{R}$ represents the set of real numbers; $A\in \mathbb{R}^{n\times m}$ is an $n\times m$ matrix with real values; $\|z\|_{A}:=z^{T}A^{-1}z$; $\|A\|_{i2}$ is the matrix induced two norm for matrix $A$; $\{x_{k}\}_{a}^{b}=:\{x_{a},x_{a+1},\cdots,x_{b}\}$ represents a sequence of vectors and $\mathcal{T}_{S}^{T}=:\{S,S+1,\cdots,T-1\}$ is the set of integers from $S$ to $T-1$. When $S=0$ the subscript is not included.  Optimal decision variables are denoted as $\hat{x}_{a|b}$ which stands for the optimal solution for $x_{a}$ at time $b$ and optimal objectives are denoted by $\widehat{J}$
\end{section}

\begin{section}{The Full Information Estimation Problem for Descriptor Systems}\label{kalman}
In this section, we provide essential introductory material that will also serve as an extension to the study in \cite{rao2001constrained} to linear causal descriptor systems.  We first present here the full information estimator (FIE) which value lies in defining a desirable state estimator.

The full information filtering problem for linear, discrete, time invariant, causal descriptor systems is defined as finding an estimate of the current state $x_{T}$ at time $T$ given the initial state estimate $\bar{x}_{0}$ and its corresponding uncertainty matrix $P_{0}$, the noisy measurement sequence $y_{1}, \cdots, y_{T}$, the input sequence $u_{0},\cdots, u_{T}$ and the following descriptor model:
\begin{align}
\label{dtss}
Ex_{k+1}&=Ax_{k}+Bu_{k}+w_{k} \\
\label{meas}
y_{k+1}&=Hx_{k+1}+v_{k}\\
x_{k} & \in \mathbb{X},~ w_{k}\in \mathbb{W},~v_{k}\in \mathbb{V}, \mbox{ for } k\in \mathcal{T}^{T} \label{constraints}
\end{align}
where $\mathbb{X},\mathbb{W},\mathbb{V}$ are convex polyhedral sets with 0 in the interiors of $\mathbb{W},\mathbb{V}$, $x_{k}\in\mathbb{R}^{n}$, $y_{k}\in\mathbb{R}^{m}$, $u_{k}\in\mathbb{R}^{q}$ and the matrices $E, A\in\mathbb{R}^{n_{1}\times n}$, $B\in\mathbb{R}^{n_{1}\times q}$ and $H\in\mathbb{R}^{m\times n}$.

\begin{subsection}{Main Assumptions}
\begin{enumerate}
\item The matrix $[E~A]$ is assumed to be full row rank; i.e. there is no dependency between the rows of the matrix pencil $\lambda E-A$.  This will eliminate the possibility of having over-determined subsystem blocks in the Kronecker canonical decomposition of the descriptor system that will constrain the input sequences $u_{k},w_{k}$ and the initial condition $x_{0}$ \cite{gantmakher1959}, \cite{brull2007}. 
\item The matrix $[E^{T}~H^{T}]^{T}$ is assumed to be full column rank.  This condition will guarantee the ability of estimating states that are unspecified by the system dynamics of \eqref{dtss}.  More precisely, if the descriptor system \eqref{dtss} contains an under-determined system block in the Kronecker decomposition, then the above condition will guarantee having measurements available for the unspecified states.  These states will be estimated using information coming from the noisy measurements only.  Observable states (states with both measurements and system dynamics) can be revealed using Kalman decomposition of descriptor systems given in \cite{banaszuk1992}.  A program was developed for this purpose and can be downloaded from \cite{kaldecomp}.
\item The index of the regular part of the descriptor system is assumed to be at most one.  In other words, the state trajectory for $x_{k}$ is independent of future values of the input and noise sequences; i.e. the system is causal.  Causality is not a limiting assumption for most practical purposes because non-causal (high index) time invariant systems can be transformed to causal descriptor systems using index reduction techniques as described for example in \cite{Luenberger1977} and \cite{Nikoukhah1999}. 
\end{enumerate}
Additional assumptions for ensuring MHE estimator stability will be given later in Appendix B.  
\end{subsection}

\begin{subsection}{Descriptor Full Information Estimator}
\begin{defn}\textbf{Full Information Estimator: \cite{Rao2000}, \cite{Rawlings2009}}\\
The full information state estimate $\hat{x}_{T|T}^{full}$ is found from solving the following minimization problem:
\begin{align}\label{full0}
\widehat{J}_{T}^{full}&:=\min_{\zeta} J_{T}(\zeta) \\
\mbox{s.t. }&x_{k}\in \mathbb{X},~ w_{k}\in \mathbb{W},~v_{k}\in \mathbb{V} \mbox{ and \eqref{dtss}-\eqref{meas}}, ~k\in\mathcal{T}^{T} \nonumber
\end{align}
where
\begin{align*}
\zeta&=\left\lbrace \{x_{k}\}_{1}^{T},\{w_{k}\}_{0}^{T-1},\{v_{k}\}_{0}^{T-1}\right\rbrace \nonumber \\
J_{T}(\zeta)&= \nonumber \\
&\left(\|Ex_{1}-A\bar{x}_{0}-Bu_{0}\|_{P_{0}^{(-)}}^{2}+\sum_{k=1}^{T-1}\|w_{k}\|_{Q}^{2}+\sum_{k=1}^{T-1}\|v_{k}\|_{R}^{2}\right) \nonumber \\ 
P_{0}^{(-)}&=AP_{0}A^{T}+Q
\end{align*}
and $\hat{x}_{T|T}^{full}$ is extracted from the $x_{T}$ element of the solution $\hat{\zeta}$ 
\end{defn}
The positive definite weighting matrices $P_{0}^{(-)},Q,R$ are specified by the user to penalize deviations according to the uncertainties as usual.  

The constrained full information estimator acquires the desirable properties of stability and optimality provided that the system is detectable and stabilizable (defined in Appendix B).  Optimality of the estimate can be established by relating to stochastic arguments assuming truncated normal distributions on $\bar{x}_{0},w_{k}$ and $v_{k}$, and finding the resulting MAP estimate.  This was shown, for example, for state space systems in \cite{Goodwin2005} and for causal linear descriptor systems recently in \cite{Almatouq2014} with the utility of quasi-Kronecker-Canonical decomposition using real transformation matrices \cite{gantmakher1959}, \cite{berger2012}.   

However, the FIE is computationally intractable since the minimization problem grows unbounded with time.  Hence, it is desired to come as close as possible to the optimality performance of FIE while using a technique that is computationally tractable.  A well known approximation technique to FIE is fixed window size MHE which will be briefly reviewed next.

\end{subsection}
\end{section}
\begin{section}{Descriptor Moving Horizon Estimation}
The moving horizon estimate is found by limiting the estimation problem in \eqref{full0} to a window of measurements and dynamic updates that slides with time while accounting for past measurements through an extra penalty cost term, referred to as an arrival cost \cite{Rao2000}.   

The arrival cost $Z_{T}(x_{T})$ for the constrained full information problem is defined by the following partial minimization problem: \cite{Rawlings2009}
\begin{align*} 
Z_{T}(x_{T}) & : =\min_{\zeta}J_{T}(\zeta)\\
\mbox{s.t. }~ & x_{k}\in \mathbb{X},~ w_{k}\in \mathbb{W},~v_{k}\in \mathbb{V}\mbox{ and \eqref{dtss}-\eqref{meas}}, ~k\in \mathcal{T}^{T-1}\nonumber 
\end{align*}
where
\[
\zeta=\left\lbrace\{x_{k}\}_{1}^{T-1},\{w_{k}\}_{0}^{T-2},\{v_{k}\}_{0}^{T-2}\right\rbrace.
\]

Hence, the constrained full information filtering problem \eqref{full0} can be rewritten as follows:
\begin{align*}
\widehat{J}_{T}^{full}& =\min_{\zeta} Z_{T-N}(x_{T-N})+\sum_{k=T-N}^{T-1}\|w_{k}\|_{Q}^{2}+\sum_{k=T-N}^{T-1}\|v_{k}\|_{R}^{2} \\
\mbox{s.t. }~& x_{k}\in \mathbb{X},~ w_{k}\in \mathbb{W},~v_{k}\in  \mathbb{V} \mbox{ and \eqref{dtss}-\eqref{meas}}, ~k\in \mathcal{T}_{T-N}^{T}\nonumber 
\end{align*}
where
\[
\zeta=\left\lbrace\{x_{k}\}_{T-N}^{T},\{w_{k}\}_{T-N}^{T-1},\{v_{k}\}_{T-N}^{T-1}\right\rbrace.
\]

Figure \ref{fig2} depicts the information coverage of the arrival cost and the sliding window cost in MHE.  Approximation techniques for finding the arrival cost, however, are inevitable as finding the exact arrival cost $Z_{T}(x_{T-N})$ analytically is a combinatorial problem \cite{faisca2008multi}.  

\input{fig2}

A technique for retaining tractability of this minimization problem is moving horizon estimation.
\begin{defn}\textbf{Moving horizon estimation\cite{Rao2000},\cite{Rawlings2009}}\\
The moving horizon state estimate $\hat{x}_{T|T}^{mhe}$ is found from solving the following minimization problem:
\begin{align}
\widehat{J}_{T}^{mh}&:=\min_{\zeta}J_{T}^{mh}(\zeta) \label{mhe1} \\
\mbox{s.t. } & x_{k}\in \mathbb{X},~ w_{k}\in \mathbb{W},~v_{k}\in \mathbb{V} \mbox{ and \eqref{dtss}-\eqref{meas}} ~~k\in \mathcal{T}_{T-N}^{T} \nonumber \\
\end{align}
where
\begin{align*}
\zeta&=\left\lbrace \{x_{k}\}_{T-N}^{T}, \{w_{k}\}_{T-N}^{T-1}, \{v_{k}\}_{T-N}^{T-1} \right\rbrace\\
J_{T}^{mh}(\zeta)&=
\bar{Z}_{T-N}^{mh}(x_{T-N})+\sum_{k=T-N}^{T-1}\|w_{k}\|_{Q}^{2}+\sum_{k=T-N}^{T-1}\|v_{k}\|_{R}^{2} 
\end{align*}
and $\hat{x}_{T|T}^{mhe}$ is extracted from the $x_{T}$ element of the solution $\hat{\zeta}$
\end{defn}
The symbol $\widehat{J}_{T}^{mh}$ is used to denote the optimum value of \eqref{mhe1}.  Here, $\bar{Z}_{T-N}^{mh}(x_{T-N})$ serves as an approximation of the arrival cost $Z_{T-N}(x_{T-N})$ and $N$ is the length of the sliding window.  One known technique for approximating $Z_{T-N}(x_{T-N})$ is using the arrival cost for the unconstrained estimation problem.

\begin{thm} \label{arrivalcost}
The arrival cost for the unconstrained full information problem at time $k=T$ is given by:
\begin{align}
J_{T}^{-}(x_{T})=&\min_{\{x_{k}\}_{1}^{T-1}}J_{T}(x_{1},x_{2},\cdots , x_{T})\nonumber \\
=&\|Ex_{T}-z_{T}\|_{P_{T-1}^{(-)}}^{2}+\widehat{J}_{T-1}\label{arrival} 
\end{align}
where, $z_{T}:=A\hat{x}_{T-1}^{(+)}+B u_{T-1}$ and $\hat{x}_{T-1}^{(+)},P_{T-1}^{(-)}$ are found from the following recursions starting at time $k=1$ and ending at time $k=T-1$:
\begin{align}\label{recur}
\hat{x}_{k}^{(+)}=&P_{k}^{(+)}H^{T}R^{-1}y_{k}+P_{k}^{(+)}E^{T}(P_{k-1}^{(-)})^{-1}(A\hat{x}_{k-1}^{(+)}+B
u_{k-1}) \nonumber \\
P_{k}^{(+)}=&(E^{T}(P_{k-1}^{(-)})^{-1}E+H^{T}R^{-1}H)^{-1} \\
P_{k}^{(-)}=&A P_{k}^{(+)}A^{T}+Q \nonumber 
\end{align}
\end{thm}
\begin{pf}
See Appendix A for detailed calculations of the arrival cost and a simple derivation for the associated descriptor Kalman recursions.  Descriptor Kalman recursions were previously derived in \cite{Darouach1993} and \cite{Ishihara2005} with the most general case derived in \cite{Nikoukhah1992}.  Descriptor Kalman recursions for the case when mixed stochastic and deterministic components are considered can be found in \cite{Almatouq2013}.  
\end{pf}

Note that the constant term $\widehat{J}_{T-1}$ in \eqref{arrival} has no influence on the minimization problem and can be eliminated.  Hence, the following prior weighting will be selected as an approximation of the true arrival cost $Z_{T-N}(x_{T-N})$:
\begin{align}\label{arrivalsel}
&\bar{Z}_{T-N}^{mh}(x_{T-N}):=\nonumber \\
&~~\|Ex_{T-N}-A\hat{x}_{T-N-1}^{(+)}-Bu_{T-N-1}\|^{2}_{P_{T-N-1}^{(-)}}
\end{align}
This selection will correspond to the exact arrival cost (minus the constant term $\widehat{J}_{T-N-1}$) when no constraints are active before time $k=T-N$.  Furthermore, this selection of the prior weighting will guarantee convergence and stability of the Moving Horizon Estimator \eqref{mhe1} as an observer as given by the following theorem.  

\begin{thm}\label{thm_mhe}
If system \eqref{dtss} with measurement sequence \eqref{meas} is detectable and stabilizable (defined in appendix B), and the noise and disturbance sequences assumed zero; i.e. $w_{k}=v_{k}=0$, then the iterative minimization of \eqref{mhe1} using prior weighting \eqref{arrivalsel} leads to convergence of the optimal state estimates to the true value of the states.  Furthermore, the resulting moving horizon estimator is an asymptotically stable observer.
\end{thm}

\begin{pf}
See appendix B. 
\end{pf}

Without loss of generality, the polyhedral constraints \eqref{constraints} can be described as inequalities in the following form:
\begin{align}\label{ineq}
E_{c} x_{k+1}\le A_{c}x_{k}+d_{c},~k \in \mathcal{T}_{T-N}^{T}
\end{align}
where $E_{c}, A_{c} \in \mathbf{R}^{n_{ineq}\times n}$ and $d_{c}\in \mathbf{R}^{n_{ineq}}$.  From now on, the decision variables for minimization will be $\{x_{k}\}_{T-N}^{T}$ alone which is possible due to the assumption $[E^{T}~H^{T}]^{T}$ that guarantees one to one correspondence between $\{\hat{x}_{k}\}_{T-N}^{T}$ and $\{\hat{w}_{k},\hat{v}_{k}\}_{T-N-1}^{T}$.  Algorithm I summarizes conventional Moving Horizon Estimation for descriptor systems.

\begin{algorithm}[t]{\textbf{Algorithm I: Descriptor System MHE}}\\
Initialization: Given $\bar{x}_{0},P_{0}$ solve \eqref{full0} up to time $T=N$\\
Find $\hat{x}_{T-N-1}^{(+)},P_{T-N-1}^{(-)}$ using \eqref{recur} with $\hat{x}_{0}^{(+)}=\bar{x}_{0},~P_{0}^{(+)}=P_{0}$\\
For $T=N,N+1,\cdots, T_{final}$:
\begin{enumerate}
\item Solve the following minimization and extract $\hat{x}_{T|T}^{mhe}$:
\begin{align*}
\min_{\{x_{k}\}_{T-N}^{T}} & \bar{Z}_{T-N}^{mh}(x_{T-N})+\sum_{k=T-N}^{T-1}\|w_{k}\|_{Q}^{2}+\sum_{k=T-N}^{T-1}\|v_{k}\|_{R}^{2} \\
\mbox{s.t. } & \mbox{\eqref{dtss}-\eqref{meas} and} \\
&E_{c}x_{k+1}\le A_{c} x_{k} +d_{c},~k\in \mathcal{T}_{T-N}^{T} 
\end{align*}
\item \textit{Update arrival cost:}
\begin{enumerate}
\item Find $\hat{x}_{T-N}^{(+)},~P_{T-N}^{(-)}$ using \eqref{recur} and $\hat{x}_{T-N-1}^{(+)},P_{T-N-1}^{(-)}$
\item Set $T=T+1$
\item Construct new arrival cost using \eqref{arrivalsel}
\end{enumerate}
\end{enumerate}
repeat
\end{algorithm}
\end{section}

\begin{section}{Multiple Window Moving Horizon Estimation}
Long horizon lengths in MHE are desirable to reach the performance limits of the full information estimator \eqref{full0}.  However, in the conventional MHE technique the problem complexity scales at least linearly with the horizon length selected. For example, excluding arrival cost calculations \eqref{arrivalsel}, an efficient interior point method implementation that exploits structure for solving \eqref{mhe1} will have a complexity of $\mathit{O}(Nn^{3})$ per Newton iteration. \cite{dunn1989efficient} Furthermore, during periods when constraints are inactive, the MHE technique conducts inequality constrained minimizations that can result in unnecessary numerical errors compared to recursive solutions for unconstrained minimizations.

In this section we develop a new strategy for moving horizon estimation for general linear descriptor systems that enables long horizons with reduced computation compared to traditional fixed window size MHE.  A small sliding window objective function is augmented with fixed cost terms in the past corresponding to states that were determined to be constrained inside the sliding window, while intermediate inequality constraints between the sliding window cost and the fixed costs are eliminated from subsequent minimizations.  This allows reformulating the objective into a significantly smaller minimization problem, especially when periods of constraint inactivity dominate.  The fixed cost terms remain in the estimation problem until there influence on the current state is negligible.  Stability is maintained by using the arrival cost for unconstrained minimization and the horizon length for the fixed cost windows are selected based on the magnitude of coupling between past and current states.  

\begin{subsection}{MW-MHE Approximation}
To begin describing the new MW-MHE approximation, the full information estimator problem \eqref{full0} is first reformulated in a form suitable for our purpose.
\begin{thm}\label{totalreform}
Given positive definite weighting matrices $P_{0}^{(-)}, R$ and $Q$ and $[E^{T}~H^{T}]^{T}$ full column rank, the full information estimator objective \eqref{full0} can be rewritten as:
\begin{align}\label{reformulate1}
\min_{\{x_{k}\}_{1}^{T}} & \sum_{k=1}^{T-N-1}\|x_{k}-\hat{x}_{k}^{sm}(x_{k+1})\|_{\Gamma_{k}^{sm}}^{2}+SC(\{x_{k}\}_{T-N}^{T}) \\
\mbox{s.t. }&E_{c}x_{k}\le A_{c} x_{k-1} +d_{c},~\eqref{dtss},\eqref{meas}~\mbox{for }k\in \mathcal{T}^{T},\nonumber
\end{align}
where
\[
SC(\{x_{k}\}_{T-N}^{T})=\bar{Z}_{T-N}^{mh}(x_{T-N})+\sum_{k=T-N}^{T-1}\left(\|w_{k}\|_{Q}^{2}+\|v_{k}\|_{R}^{2}\right) 
\]
and
\begin{align}
\hat{x}_{k}^{sm}(x_{k+1})&=\hat{x}_{k}^{(+)}+\Gamma_{k}^{sm}A^{T}Q^{-1}(Ex_{k+1}-A\hat{x}_{k}^{(+)}-Bu_{k}) \nonumber \\
\Gamma_{k}^{sm}&=((P_{k}^{(+)})^{-1}+A^{T}Q^{-1}A)^{-1} \label{smooth2}
\end{align}
\end{thm}
\begin{pf}
The proof is by induction; using repeated measurement and time update reformulations for $k=1,\cdots,T$, as described in Appendix A, the cumulating objective will result in \eqref{reformulate1}.
\end{pf}

\begin{rem}
The recursions for $\hat{x}_{k}^{sm}(x_{k+1}),\Gamma_{k}^{sm}$ presented in \eqref{smooth2} correspond to the Kalman smoothing recursions for descriptor systems derived in Appendix A and also in \cite{Ishihara2005}.  Note that $\hat{x}_{k}^{sm}$ depends on the decision variable $x_{k+1}$.  
\end{rem}

\input{fig3}

Referring to Figure \ref{fig3}, suppose we desire to approximate the minimization of \eqref{reformulate1} by dropping the inequality constraints \eqref{ineq} before time $k=a_{s}$ and within the time interval $k\in[b_{s}+1,a_{(s+1)}-1]$ after assuming constraint inactivity in this region, where $a_{s} < b_{s} < a_{(s+1)} < T$.  Consequently, the objective \eqref{reformulate1} can be partitioned as follows:
\begin{align}
&J_{T}^{mw}:= \overbrace{\sum_{k=1}^{a_{s}-1}\|x_{k}-\hat{x}_{k}^{sm}(x_{k+1})\|_{\Gamma_{k}^{sm}}^{2}}^{:=IC(\{x_{k}\}_{1}^{a_{s}-1})}+\overbrace{\sum_{k=a_{s}}^{b_{s}}\|x_{k}-\hat{x}_{k}^{sm}(x_{k+1})\|_{\Gamma_{k}^{sm}}^{2}}^{:=FC_{s}(\{x_{k}\}_{a_{s}}^{b_{s}})}\nonumber \\
&+\overbrace{\sum_{k=b_{s}+1}^{a_{(s+1)}-1}\|x_{k}-\hat{x}_{k}^{sm}(x_{k+1})\|_{\Gamma_{k}^{sm}}^{2}}^{:=UC_{s}(\{x_{k}\}_{b_{s}+1}^{a_{(s+1)}-1})}+SC(\{x_{k}\}_{T-N}^{T})\label{mhe3}
\end{align}
where $J_{T}^{mw}$ denotes the MW-MHE objective, $IC(\{x_{k}\}_{1}^{a_{s}-1})$ an initial cost window with no constrained arguments, $FC_{s}(\{x_{k}\}_{a_{s}}^{b_{s}})$ a fixed cost window with constrained arguments and $UC_{s}(\{x_{k}\}_{b_{s}+1}^{a_{(s+1)}-1})$ an unconstrained fixed cost window.  The subscripts on $FC_{s},UC_{s}$ is used to allow multiple fixed cost windows as introduced later.  Since inequality constraints are not imposed before time $a_{s}$ clearly we have $IC=0$ by selecting $x_{k}=\hat{x}_{k}^{sm}(x_{k+1}),~k=1,2,\cdots,a_{s}-1$.  Also, since no inequality constraints are imposed on the intermediate cost $UC_{s}$, we can partially minimize this term as:\\
$UC_{s}^{(-)}(x_{b_{s}+1},x_{a_{(s+1)}-1}):=\min_{\{x_{k}\}_{b_{s}+2}^{a_{(s+1)}-2}}UC_{s}(\{x_{k}\}_{b_{s}+1}^{a_{(s+1)}-1}) \\
=\|x_{b_{s}+1}-\hat{x}_{b_{s}+1}^{sm}\|_{\Gamma_{b_{s}+1}^{sm}}^{2}+\|x_{a_{(s+1)}-1}-\hat{x}_{a_{(s+1)}-1}^{sm}\|_{\Gamma_{a_{(s+1)}-1}^{sm}}^{2}$\\
Furthermore, by subsequent application of the recursions in \eqref{smooth2}, we may express $\hat{x}_{b_{s}+1}^{sm}$ in terms of $x_{a_{(s+1)}-1}$ as follows:
\begin{align}
\hat{x}_{b_{s}+1}^{sm}=&\hat{x}_{b_{s}+1}^{(+)}+\Gamma_{b_{s}+1}^{sm}
A^{T}Q^{-1}(Ex_{b_{s}+2}-A\hat{x}_{b_{s}+ 1}^{(+)}-Bu_{b_{s}+1})\nonumber \\
=&M_{2}^{s}x_{b_{s}+2}+r_{1}^{s}\nonumber \\
=&M_{3}^{s}x_{b_{s}+3}+M_{2}^{s}r_{2}^{s}+r_{1}^{s}=\cdots  \nonumber \\
=&M_{c_{s}}^{s}x_{a_{(s+1)}-1}+
\sum_{i=1}^{c_{s}-1}M_{i}^{s}r_{i}^{s} \label{xtau2}
\end{align}
where
\begin{align}
M_{1}^{s}=&I,~~M_{q}^{s}=M_{q-1}^{s}\Gamma_{b_{s}+q-1}^{sm}A^{T}Q^{-1}E,\label{Mi} \\
r_{q-1}^{s}=&\hat{x}_{b_{s}+q-1}^{(+)}-\Gamma_{b_{s}+q-1}^{sm}A^{T}Q^{-1}(A\hat{x}_{b_{s}+q-1}^{(+)}+Bu_{b_{s}+q-1}) \nonumber \\
c_{s}=&a_{(s+1)}-b_{s}, q=2,3,\cdots, c_{s} \label{ri}
\end{align}
Thus we may rewrite the intermediate unconstrained cost as:
\begin{align}
UC_{s}^{(-)}=&\|x_{b_{s}+1}-M_{c_{s}}^{s}x_{a_{(s+1)}-1}-
\sum_{i=1}^{c_{s}-1}M_{i}^{s}r_{i}^{s}\|_{\Gamma_{b_{s}+1}^{sm}}^{2} \nonumber \\
+&~~~\|x_{a_{(s+1)}-1}-\hat{x}_{a_{(s+1)}-1}^{sm}(x_{T-N})\|_{\Gamma_{a_{(s+1)}-1}^{sm}}^{2} \label{uncon}
\end{align}
where $M_{i}^{s},r_{i}^{s}$ are given by the recursions in \eqref{Mi}, \eqref{ri}.  Consequently, the MW-MHE problem \eqref{mhe3} becomes:
\begin{align}\label{simpmhe}
\widehat{J}_{T}^{mw}&=\min_{\zeta} ~FC_{s}(\{x_{k}\}_{a_{s}}^{b_{s}})+\nonumber \\
&~\hspace{.5in}UC_{s}^{(-)}(x_{b_{s}+1},x_{a_{(s+1)}-1})+SC(\{x_{k}\}_{T-N}^{T})\\
\mbox{s.t. } & E_{c}x_{k+1}\le A_{c}x_{k}+d_{c},~ k\in\{[a_{s},b_{s}]\cup \mathcal{T}_{T-N}^{T}\} \nonumber 
\end{align}
where
\[
\zeta=\left\lbrace \{x_{k}\}_{a_{s}-1}^{b_{s}+1}, \{x_{k}\}_{a_{(s+1)}-1}^{T}\right\rbrace.
\]

This is a convenient form since the intermediate unconstrained states $x_{b_{s}+2},\cdots, x_{a_{(s+1)}-2}$ are eliminated from the objective function \eqref{mhe3} at the expense of simple recursive calculations for finding $M_{i}^{s}$, $r_{i}^{s}$ given by \eqref{Mi}, \eqref{ri}.  Excluding the cost for calculating $\Gamma_{k}^{sm}$, an efficient interior point method implementation that exploits structure for solving \eqref{simpmhe} will have an approximate complexity of $\sim \mathit{O}((T-N+1)n^3)+\mathit{O}((b_{s}-a_{s}+3)n^{3})$ per Newton iteration.  Hence, when the intermediate interval $[b_{s}+1,a_{(s+1)}-1]$ is large and assumed to be a region where no constraints are active then significant complexity reductions for long horizon estimation problems can be achieved.  The extra calculations involved for finding $\Gamma_{k}^{sm}$ per iteration are of order $\sim \mathit{O}(n^3)$ that can be made efficient using square root factors. \cite{chisci1992}  

\input{fig4}

\end{subsection}

\begin{subsection}{MW-MHE Algorithm}
Based on the MW-MHE approximation, a general algorithm can be synthesized that can handle multiple fixed and unconstrained windows.  This is the basis of the new MW-MHE algorithm that is depicted in the flowchart given in Figure \ref{fig4}.  An explanation of the algorithm will proceed.  

Starting at time $T=1$, for the first $N$ iterations, we solve the full information estimator problem given in \eqref{full0}.  At each subsequent time; i.e. $T>N$, the following MW-MHE minimization problem is solved:
\begin{align} \label{mwmhe}
\min_{\zeta} & \sum_{s=s_{min}}^{s}\left(FC_{s}(\{x_{k}\}_{a_{s}}^{b_{s}})+UC_{s}^{(-)}(x_{b_{s}+1},x_{a_{(s+1)}-1})\right)\nonumber \\
&~\hspace{1.5in}+SC(\{x_{k}\}_{T-N}^{T}) \\
\mbox{s.t. }& E_{c}x_{k+1}\le A_{c}x_{k}+d_{c},~ k\in\{\{[a_{s},b_{s}]\}_{s_{min}}^{S}\cup \mathcal{T}_{T-N}^{T}\},\nonumber 
\end{align}
where
\[
\zeta=\left\lbrace \{x_{k}\}_{a_{s}}^{b_{s}+1},~x_{a_{(s+1)}-1},~\{x_{k}\}_{T-N}^{T}\right\rbrace. 
\]
Note that the number of fixed cost windows is indexed by $s$; $s=s_{min}$ is the index of the first fixed cost window (farthest in time), $s=S$ is the index of the last fixed cost window (nearest in time).  For each fixed cost window $s$, the time when constraints first became active after exiting the sliding window is recorded in $a_{s}$ and the subsequent time when constraints first became inactive (after being active) is recorded in $b_{s}$.  Consequently, $FC_{s}(\{x_{k}\}_{a_{s}}^{b_{s}})$ is the $s$th fixed cost window associated with the interval $[a_{s},b_{s}]$ and $UC_{s}^{(-)}(x_{b_{s}+1},x_{a_{(s+1)}-1})$ is the $s$th unconstrained cost window associated with the interval $[b_{s}+1,a_{(s+1)}-1]$. Note that the first term in $FC_{s_{min}}(\{x_{k}\}_{a^{s_{min}}}^{b^{s_{min}}})$ corresponds to the arrival cost of the unconstrained minimization upto time $a^{s_{min}}-1$.  The filtered estimate $\hat{x}_{T|T}^{mw}$ and the smooth estimate $\hat{x}_{T-N|T}^{mw}$ are extracted from the minimizer of \eqref{mwmhe} $\hat{\zeta}$ after every minimization.  At each iteration, the index set of active constraints $\mathcal{A}(T-N)$ is found as follows: 
\begin{align}
\mathcal{A}(T-N):=\{l:E_{c}^{l}\hat{x}_{T-N|T}^{mw} = A_{c}^{l}\hat{x}_{T-N-1|T}^{sm}+d_{c}^{l}\} \label{actset}
\end{align}
where $E_{c}^{l}, A_{c}^{l}, d_{c}^{l}$ correspond to the $l$th row of $E_{c},A_{c},d_{c}$ respectively and $\hat{x}_{T-N-1|T}^{sm}$ is the corresponding unconstrained smooth estimate found from \eqref{smooth2} with $x_{T-N}=\hat{x}_{T-N|T}^{mw}$.  The active constraint set can be deduced, for example, using the dual variables in primal-dual interior point solvers.  Beginning with time $T=N$, if the active constraint set $\mathcal{A}(T-N)$ is empty, then no fixed cost or unconstrained cost windows are formed and normal moving horizon estimation proceeds by minimizing $SC(\{x_{k}\}_{T-N}^{T})$.  If, however, the active set $\mathcal{A}(T-N)$ is non-empty, then a new fixed cost window, indexed by $i=S$ is constructed by setting $a_{s}=b_{s}=T-N$.  The index $b_{s}$ is incremented if the active set $\mathcal{A}(T-N)$ continues to be non-empty after subsequent minimizations.  At the same time $FC_{s}(\{x_{k}\}_{a_{s}}^{b_{s}})$ is constructed as follows:
\begin{align}\label{FC}
FC_{s}(\{x_{k}\}_{a_{s}}^{b_{s}}) = \sum_{k=a_{s}}^{b_{s}}\|x_{k}-\hat{x}_{k}^{sm}(x_{k+1})\|_{\Gamma_{k}^{sm}}^{2}
\end{align}  
At the first point in time when $\mathcal{A}(T-N)$ becomes empty (after being non-empty), the fixed cost window is "detached" in that no further elements are added and an unconstrained fixed cost window is constructed by recording $a_{(s+1)}=T-N$ and setting a "New Window" flag to "on".  The timer $a_{(s+1)}$ is incremented until $\mathcal{A}(T-N)$ becomes non-empty again (after being empty).  At the same time the recursions for expressing $\hat{x}_{b_{s}+1}^{sm}(x_{b_{s}+2})$ in terms of $\hat{x}_{a_{(s+1)}-1}^{sm}(x_{T-N})$  are updated using \eqref{Mi}, \eqref{ri} to construct $UC_{s}^{(-)}(x_{b_{s}+1},x_{a_{(s+1)}-1})$ according to \eqref{uncon}.

At any point in time there will be $S-s_{min}+1$ fixed cost windows and $S-s_{min}+1$ unconstrained cost windows that are retrieved from memory to construct the objective function \eqref{mwmhe}.  This objective is minimized subject to the inequality constraints \eqref{ineq} within the time interval specified by the time indices $a_{s},b_{s}$ and within the sliding window interval $[T-N,T]$.  If at any time the condition $T>b^{s_{min}}+N+N_{FC}+1$ is satisfied, where $N_{FC}$ is a tuning parameter to be defined later, then both the fixed cost window and the unconstrained cost window furthest in time, indexed by $s_{min}$, are eliminated from the minimization problem \eqref{mwmhe} and the new objective is constructed accordingly for the next minimization. 

Using this algorithm, a significant reduction in problem size complexity for long horizon length estimation problems is possible with guaranteed stability and with less numerical errors.  Moreover, if the assumption of inactive constraints within the unconstrained regions is satisfied, then the performance of the estimator will approach that of the FIE given by solving \eqref{full0}.  This will be demonstrated in the next section with an example.
\end{subsection}

\begin{subsection}{Horizon Length Selection}
The new MW-MHE algorithm promotes selecting horizon lengths based on the sensitivity between remote states in time rather than based on implementation restrictions.  This new criteria for selecting the horizon length will now be developed.  The magnitude of coupling between terminal states $x_{b_{s}+1}$ and $x_{a_{(s+1)}-1}$, can be inferred from the first cost term in \eqref{uncon} via the following matrix norm:
\begin{align}\label{coupling}
\|(\Gamma_{b_{s}+1}^{sm})^{-1}M_{c_{s}}^{s}\|_{i2}
\end{align}
where $\|\cdot\|_{i2}$ corresponds to the matrix induced 2-norm.  We establish the following stability theorem.
\begin{thm}\label{condnum}
Given that the system \eqref{dtss},\eqref{meas} is detectable and stabalizable \cite{Nikoukhah1992}, then: 
\begin{align} \label{converge}
\|(\Gamma_{b_{s}+1}^{sm})^{-1}M_{c_{s}}^{s}\|_{i2} \rightarrow 0 \mbox{ as } c_{s},~b_{s} \rightarrow \infty
\end{align}
\end{thm}
\begin{pf}
See Appendix C for proof of this theorem and Appendix B for definitions of detectability and stabalizability.
\end{pf}
This theorem implies that the wider the gap in time between the terminal states $x_{b_{s}+1}$ and $x_{a_{(s+1)}-1}$, the less sensitive their values become to each other.  Using this measure of dependency, we define a maximum lag tuning parameter $N_{FC}$, which corresponds to the maximum number of sliding window minimization steps required before dropping the inequality constraints within the time interval $k\in [a_{s},b_{s}]$ that is selected based on the following criteria:
\begin{align}\label{Nfc}
\|(\Gamma_{b_{s}+1}^{sm})^{-1}M_{N_{FC}}^{s}\|_{i2} \le U
\end{align}
where $U$ is a specified upper bound on the magnitude of acceptable coupling between remote states selected by the user.  Consequently, referring again to Figure \ref{fig3} when $T>b_{s}+N_{FC}+N+1$, the inequality constraints within the interval $k\in [a_{s},b_{s}]$ can be safely dropped based on the specified maximum sensitivity between the remote states $x_{b_{s}+1}$ and $x_{a_{(s+1)}-1}$.  Upon satisfying this condition and dropping the inequality constraints, the arrival cost approximation at time $k=T-N$ will be embedded in $SC(\{x_{k}\}_{T-N}^{T})$ given by $\bar{Z}^{mh}_{T-N}(x_{T-N})$, which is the unconstrained arrival cost term used in normal MHE (Algorithm I).  A fixed value of $N_{FC}$ can be selected based on the steady state value of $\Gamma_{k}^{sm}$ which corresponds to the solution of the algebraic Riccati equation \eqref{riccati} shown in Appendix C.
\end{subsection}

\begin{rem}
The stability proof for MW-MHE follows the stability proof for MHE given in Appendix B since the reformulation \eqref{reformulate1} was used in showing convergence and stability of the MHE estimator.  It can be also argued that the MW-MHE is essentially dropping inequality constraints from the normal MHE which has no destabilizing effect since the unconstrained MHE is essentially the descriptor Kalman filter that is stable under detectability and stabilizability assumptions as given in \cite{Nikoukhah1992}.
\end{rem}

\begin{rem}
The basis for dropping inequality constraints in the MW-MHE is the assumption that if the state exits the sliding window with no constraints active, then the unconstrained solution given by \eqref{smooth2} will not violate any constraints in future minimizations.  This hypothesis can be verified, if desired, using the smooth recursions given in \eqref{smooth2}.  Another improvement to the technique is to impose only the inequality constraints identified by the active set $\mathcal{A}(T-N)$ and to drop the inequalities once the state becomes inactive after smoothing.  Nevertheless, it was observed in simulation that these added improvements have less significance than dropping inequalities for the inactive states once exiting the sliding window.
\end{rem}

\end{section}

\begin{section}{Example}
To illustrate the performance of the MW-MHE algorithm, the numerical example of the electromechanical actuator with an unknown input presented in \cite{Boulkroune2010} will be used with additional inequality constraints on the unknown input.  The state variables are the motor shaft velocity $\omega_{m}$, the elastic torque $\delta_{\theta}$ and the load shaft velocity $\omega_{c}$, while the control input is the stator current $i_{e}$.  The unknown disturbance $d$ is due to coulomb friction and load disturbances with known lower and upper limits given as $d_{l}=-35$ and $d_{u}=35$ respectively.  The objective is to estimate the state vector $x_{k}=[\omega_{m},\omega_{c},\delta_{\theta}]^{T}$ and the unknown input $d$ using moving horizon estimation.  For details of the model equations, model parameters and estimator parameters the reader is referred to \cite{Boulkroune2010}.  

The descriptor model was simulated using randomly generated disturbance sequences of zero mean and variance 1.  The value of $d$ was varied in steps as shown in the lower part of Figure \ref{fig5}.  Additive white Gaussian noise was then added to the output measurements obtained from simulation with zero mean and variance of 0.1. 

We first implemented the full information estimator (FIE) given by \eqref{full0} in Section 2.  Second, the normal moving horizon estimator (MHE) given by Algorithm I was implemented with different horizon lengths $N$.  Third, the multiple window moving horizon estimator (MW-MHE) was implemented with a horizon length of $N=1$ and different values for the time lag parameter $N_{FC}$.  The result of these experiments are shown in Table \ref{tab1}.

Table \ref{tab1} shows the comparison in terms of total mean square error performance for all the four state estimates combined.  The amount of reduction in computation time that was achieved using MW-MHE compared to MHE is also shown as a percentage.  Finally, the associated magnitude of coupling between distant states using equation \eqref{coupling} was found for each value of $N_{FC}$.

The results indicate that the full information estimator gives a lower bound on mean square error performance of $120$.  Also, the mean square error values for both the MHE and MW-MHE almost match for the horizon lengths selected as expected, but deviate with long horizons in favour for the MW-MHE algorithm.  This can be attributed to less numerical errors, which is expected since the optimization problems are smaller in size.  Also, from Table \ref{tab1}, we notice that as the horizon length increases both MHE and MW-MHE give lower mean square error values and approaches the m.s.e value for FIE.  The reduction in computation time achieved by the MW-MHE ranges from $17\%-56\%$ with more reduction at longer horizons.  The last column in Table \ref{tab1} shows that the magnitude of coupling between distant states decreases with increasing values of $N_{FC}$ as predicted in this study.  Figure \ref{fig5} shows one of the estimation results for $\omega_{m}$ and $d$ for comparison.

\begin{table*}[t]\label{tab1}
\begin{center}
\begin{tabular}{c c c c c c c}
FULL (mse) & MHE (mse) & $N$ & MW-MHE(mse) & $(N,N_{FC})$ &  time reduction $\%$ & $\|(\Gamma_{1}^{sm})^{-1}M_{N_{FC}}^{s}\|_{2}$ \\
\hline
120.1 & 319.9 & 5 & 318.8 & (1,4) & -17 $\%$ & 0.2872\\
120.1 & 215.6 & 10 & 213.0 & (1,9) & -34 $\%$ & 0.281\\
120.1 & 177.8 & 15 & 173.7 & (1,14) & -42 $\%$ & 0.27\\
120.1 & 155.6 & 20 & 150.5 & (1,19) & -48 $\%$ & 0.254\\
120.1 & 132.0 & 30 & 125.0 & (1,29) & -56 $\%$ & 0.215\\
\end{tabular}
\caption{Performance Comparison between FIE, MHE and MW-MHE for the Electromechanical Actuator Example}
\end{center}
\end{table*}

\begin{figure}[t]
\begin{center}
\includegraphics[scale=0.35]{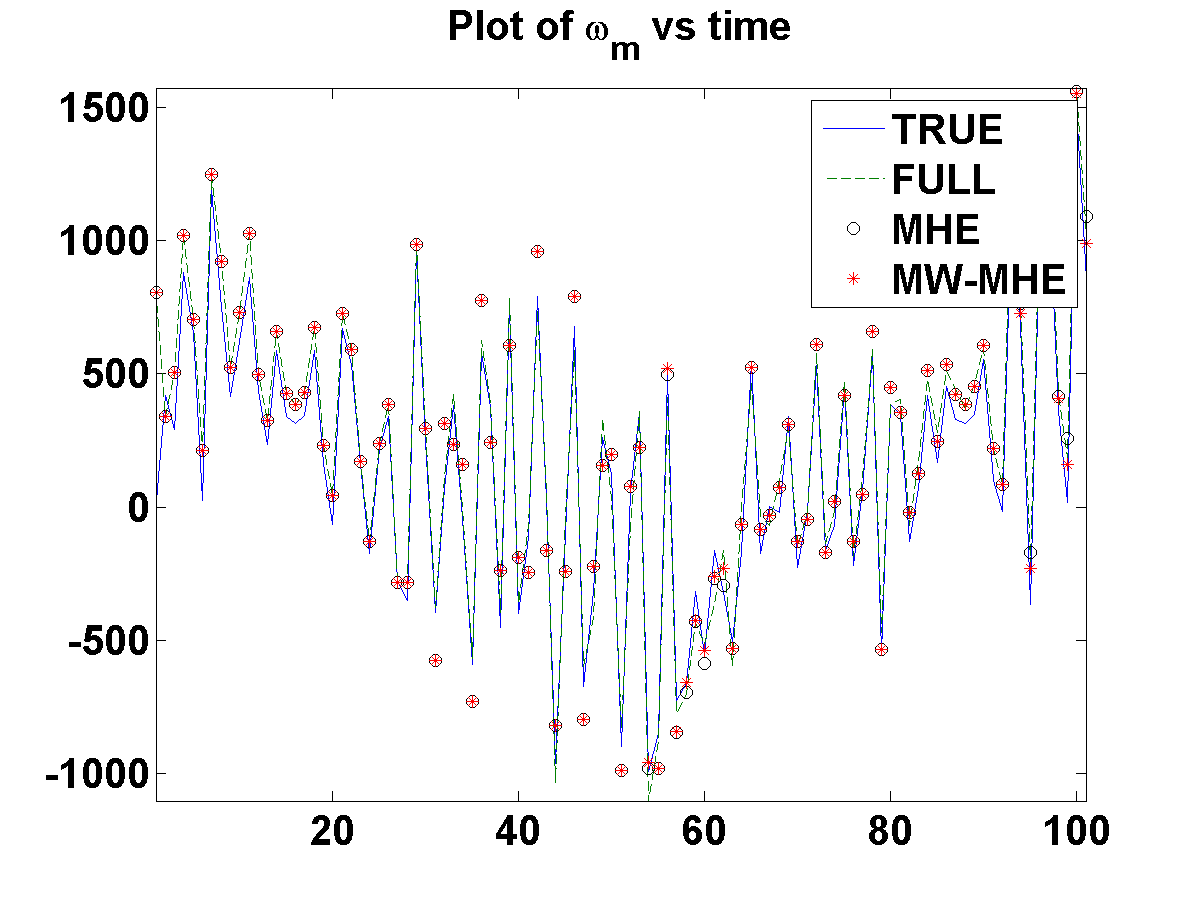}
\includegraphics[scale=0.35]{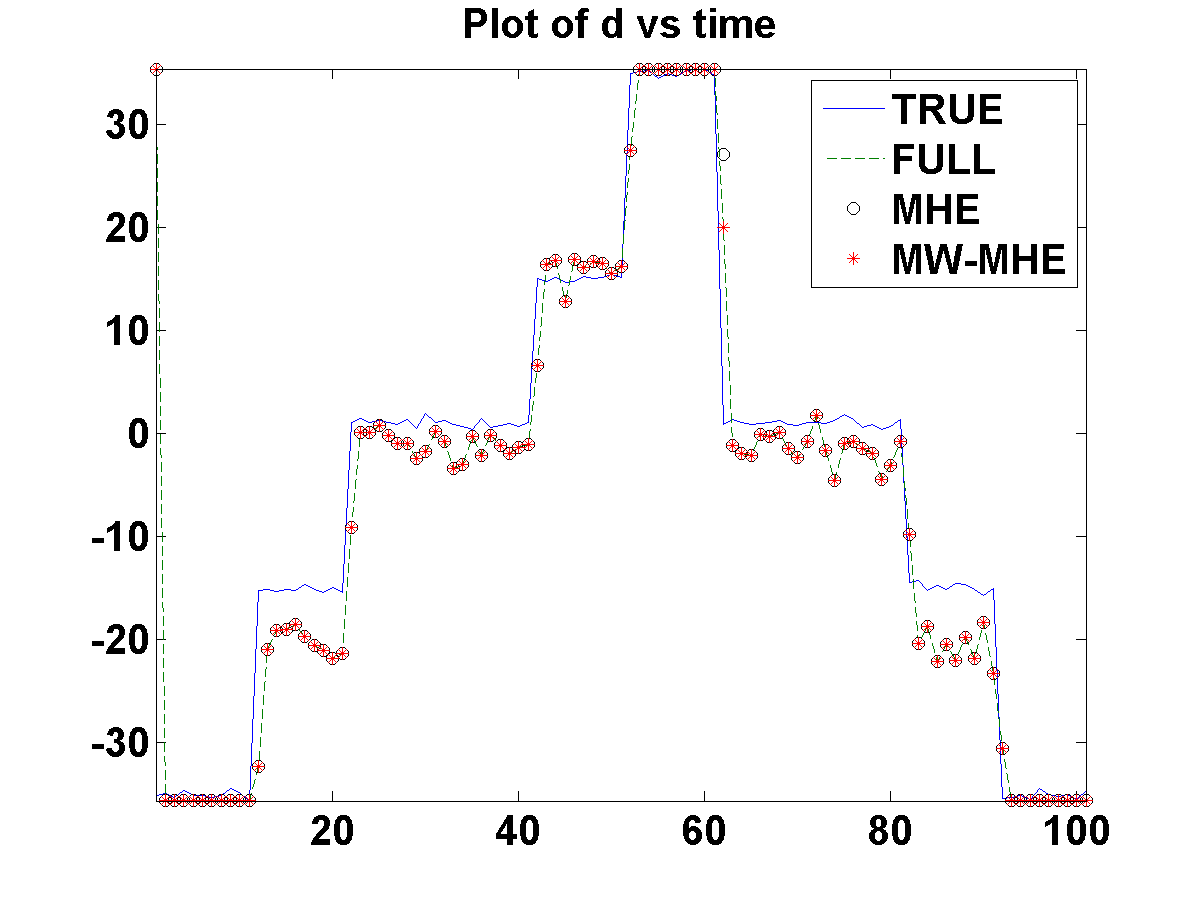} \label{fig5}
\caption{{Results for estimating $x_{1}=\omega_{m}$ (top) and $d$ (bottom) with $N=30$ and $N_{FC}=29$}}
\end{center}
\end{figure}

\end{section}

\begin{section}{Conclusion}
A new strategy for moving horizon estimation for general linear descriptor systems that enables long horizons with reduced computation compared to traditional techniques was developed.  A method for selecting the horizon length based on a condition number of a matrix that couples remote states was also developed.  Computational efficiency was achieved by exploiting constrained inactivity and numerical errors were reduced by using a short sliding window objective.  Moreover, our analysis was generalized for descriptor systems that admits estimation using differential algebraic models and problems involving unknown inputs.  Estimator stability was proven using the arrival cost for unconstrained estimation.  The example presented show the advantages using this new strategy in reducing computational requirements and numerical errors associated with long horizon estimation.

Extension to non-linear moving horizon estimation will not include simple recursions as in the linear case.  However, intelligent management of the inequality constraints can be achieved using similar techniques to the ones presented in this paper.  This and other problems involving $\ell_{1}$ and Huber penalties will be subjects for future studies.
\end{section}

\begin{ack}                               
This work was partially supported by the Saudi Arabian Ministry of Higher Eduction.
\end{ack}

\appendix
\section{\textbf{Descriptor MHE Arrival Cost Calculation}}
The following lemma will be used in finding the arrival cost for descriptor MHE.
\begin{lem}\label{dynamic} \cite{Almatouq2013} Assuming the matrices $P$ and $S$ are symmetric positive definite, the following are identities:
\begin{align}\label{ident1}
\|x-z\|^{2}_{P}+\|y-Mx\|^{2}_{s}&=\|x-\hat{x}_{1}\|^{2}_{\Gamma_{1}}+\|y-Mz\|^{2}_{\Sigma}\\
\|Ex-z\|^{2}_{P}+\|y-Mx\|^{2}_{s}&=\|x-\hat{x}_{2}\|^{2}_{\Gamma_{2}}+\|E\hat{x}_{2}-z\|^{2}_{P}\nonumber \\
&~~~~+\|y-M\hat{x}_{2}\|^{2}_{s}\label{ident2}
\end{align} where,
\begin{align*}
\Sigma&=MPM^{T}+S\\
\Gamma_{1}&=(P^{-1}+M^{T}S^{-1}M)^{-1} &\Gamma_{2}&=(E^{T}P^{-1}E+M^{T}S^{-1}M)^{-1}\\
\hat{x}_{1}&=z+\Gamma_{1} M^{T}S^{-1}(y-Mz) 
&\hat{x}_{2}&=\Gamma_{2}(E^{T}P^{-1}z+M^{T}S^{-1}y)
\end{align*}
\end{lem}
\begin{pf}
See appendix section of \cite{Almatouq2013}.
\end{pf}

\textbf{Proof of Theorem \eqref{arrivalcost}}\\
The objective function \eqref{full0} at time $k=1$ is given by:
\begin{align}\label{J1meas}
J_{1}(x_{1})&=(\overbrace{\|Ex_{1}-A\hat{x}_{0}^{(+)}-Bu_{0}\|^{2}_{P_{0}^{(-)}}+\|y_{1}-Hx_{1}\|^{2}_{R}}^{\mbox{Measurement Update 1}})
\end{align}
Using equation \eqref{ident2}, the following substitutions can be made: $x\leftarrow  x_{1},~z\leftarrow A\hat{x}_{0}^{(+)}+Bu_{0}=z_{0},~P\leftarrow P_{0}^{(-)},~y\leftarrow y_{1},~M\leftarrow H,~S\leftarrow R,~\hat{x}_{2}\leftarrow\hat{x}_{1}^{(+)}$ and $\Gamma\leftarrow P_{1}^{(+)}$.  Consequently, \eqref{J1meas} can be written as:
\begin{align*}
J_{1}(x_{1})&=\|x_{1}-\hat{x}_{1}^{(+)}\|^{2}_{P_{1}^{(+)}}+\widehat{J}_{1}\\
&\mbox{where}\\
P_{1}^{(+)}&=(E^{T}(P_{0}^{(-)})^{-1}E+H^{T}R^{-1}H)^{-1}
\nonumber \\
\hat{x}_{1}^{(+)}&=P_{1}^{(+)}(E^{T}(P_{0}^{(-)})^{-1}z_{0}+H^{T}R^{-1}y_{1}) \nonumber \\
\widehat{J}_{1}&=\|y_{1}-H\hat{x}_{1}^{(+)}\|^{2}_{R}+\|E\hat{x}_{1}^{(+)}-A\hat{x}_{0}^{(+)}-Bu_{0}\|^{2}_{P_{0}^{(-)}} 
\end{align*}
Similar reformulations can be used during any subsequent measurement updates.
The matrix $P_{1}^{(+)}$ is symmetric positive definite; i.e.
\begin{align}
x^{T}[E^{T}~H^{T}]\left[\begin{array}{cc}(P_{0}^{(-)})^{-1}&0\\0&R^{-1}\end{array}\right]\left[\begin{array}{c}E\\H\end{array}\right]x>0~~\forall x\ne 0
\end{align}
because $[E^{T}~H^{T}]^{T}$ is full column rank and $P_{0}$ and $R$ are both symmetric positive definite by assumption.

The objective \eqref{full0} at time $k=1$ with a time update can be written as:
\begin{align}\label{J1}
(\overbrace{\|x_{1}-\hat{x}_{1}^{(+)}\|^{2}_{P_{1}^{(+)}}+\|Ex_{2}-Ax_{1}-Bu_{1}\|_{Q}^{2}}^{\mbox{Time Update 1}})+\widehat{J}_{1}
\end{align}
Since $P_{1}^{(+)}$ and $Q$ are both symmetric positive definite, identity \eqref{ident1} can be used to reformulate the objective.  After making the appropriate substitutions into \eqref{ident1}, \eqref{J1} will become:
\begin{align}\label{timeupdate}
&(\|x_{1}-\hat{x}_{1}^{sm}(x_{2})\|^{2}_{\Gamma_{1}^{sm}}+\|Ex_{2}-A\hat{x}_{1}^{(+)}-Bu_{1}\|^{2}_{P_{1}^{(-)}}+\widehat{J}_{1})\\
&P_{1}^{(-)}=A P_{1}^{(+)}A^{T}+Q \nonumber \\
&\Gamma_{1}^{sm}=((P_{1}^{(+)})^{-1}+A^{T}Q^{-1}A)^{-1} \nonumber \\
&\hat{x}_{1}^{sm}(x_{2})=\hat{x}_{1}^{(+)}+\Gamma_{1}^{sm}A^{T}Q^{-1}(Ex_{2}-A\hat{x}_{1}^{(+)}-Bu_{1})\label{timeupdate2}
\end{align}
Notice, that $\hat{x}_{1}^{sm}$ depends on $x_{2}$ and hence provides a smoothed like estimate when constraints \eqref{constraints} are not taken into account.  Note also that both $P_{1}^{(-)}$ and $\Gamma_{1}^{sm}$ are both symmetric positive definite since $P_{1}^{(+)}$ and $Q$ are symmetric positive definite.

Now the arrival cost term at time $k=1$ can be obtained by minimizing \eqref{timeupdate} over $x_{1}$.  This can be achieved by selecting $x_{1}$ to be equal to \eqref{timeupdate2} to obtain the following:
\begin{align}\label{arivalcalc}
J_{2}^{-}(x_{2})=&\|Ex_{2}-z_{1}\|^{2}_{P_{1}^{(-)}}+\widehat{J}_{1}
\end{align}
where, $z_{1}:=A\hat{x}_{1}^{(+)}-B u_{1}$.  By induction, subsequent measurement and time updates for $k=2,\cdots,T-1$ will result in the arrival cost given by \eqref{arrival}.

\section{\textbf{Proof of Theorem \ref{thm_mhe}}}
The necessary detectability and stabilizability assumptions are first presented, followed by a formal definition of observer asymptotic stability.

\begin{assum}{\textit{Detectability:}\cite{Nikoukhah1992}}\\
System \eqref{dtss}, \eqref{meas} is called detectable if:
\begin{align*}
rank(\left[\begin{array}{c}\mu E-\phi A\\ H\end{array}\right])=n\\
\end{align*}
for all complex pairs $(\mu,\phi)\ne (0,0)$ such that $|\mu| \ge |\phi|$.  
\end{assum}

\begin{assum}{\textit{Stabilizability:}\cite{Nikoukhah1992}}\\
System \eqref{dtss} is called stabilizable if:
\begin{align*}
rank(\left[\begin{array}{ccc}\mu E-\phi A& Q & 0\\ \mu H & 0 & R \end{array}\right])=n_{1}+m
\end{align*}
for almost all complex pairs $(\mu,\phi)\ne (0,0)$ and $|\mu| \ge |\phi|$. 
\end{assum}
Note, verifying detectability and stabilizability using the above conditions is difficult in practice.  The alternative is to use Kalman decomposition of descriptor systems \cite{banaszuk1992},\cite{kaldecomp}. 

\begin{defn}\textbf{Observer Asymptotic Stability\cite{Rao2000}:} \label{stabdef}\\
The Moving Horizon estimator \eqref{mhe1}, which results in finding the estimates $\hat{x}_{1}^{mh},\cdots,\hat{x}_{T}^{mh}$, is an asymptotically stable observer for the system:
\begin{align}\label{dtss2}
Ex_{k+1}=&A x_{k}+B u_{k} \\
\label{meas2}
y_{k+1}=&Hx_{k+1}
\end{align}
if for any $\epsilon>0$, there corresponds a $\delta>0$ and a positive integer $\tilde{T}$ such that if $\|Ex_{1}-A\hat{x}_{0}^{(+)}-Bu_{0}\| \le \delta$
then $\|\hat{x}_{T}^{mh}-x_{T}^{*}\| \le \epsilon$  $\forall T\ge \tilde{T}$.  Furthermore, $\hat{x}_{T}^{mh} \rightarrow x_{T}^{*}$ as $T\rightarrow \infty$, where $x_{T}^{*}$ are the true values of the state found by solving \eqref{dtss2}.
\end{defn}

\textbf{Proof of Theorem \ref{thm_mhe}}\\
The presented proof is analogous to the proofs presented in \cite{Rao2000} and in \cite{Rawlings2009} for state space systems and is extended to descriptor systems.   We first make reference to stability results of the descriptor Kalman filter given in \cite{Nikoukhah1992}.  We then find the limiting value of the arrival cost \eqref{arrivalsel} based on Kalman filter convergence which allows us to find an upper bound for the moving horizon cost sequence $\{\widehat{J}_{T}^{mhe}\}$.  The limiting values of the moving horizon estimates $\hat{x}_{T}^{mh}$ are then sought followed by application of Definition \ref{stabdef}.

\begin{thm}\label{stablity} \cite{Nikoukhah1992}\\
Suppose that \eqref{dtss}-\eqref{meas} is both detectable and stabilizable, then for any initial condition $P_{0}^{(+)}>0$, the recursion for $P_{k}^{(+)}$ converges exponentially fast to $P_{\infty}^{(+)}$ which is the unique positive semi-definite solution of the algebraic descriptor Riccati equation:
\begin{align}\label{riccati1}
P_{\infty}^{(+)}=&(E^{T}(A P_{\infty}^{(+)}A^{T}+Q)^{-1}E+H^{T}R^{-1} H)^{-1}
\end{align}
Furthermore, the steady state Kalman filter given by:
\begin{align}\label{kal}
\hat{x}_{k+1}^{(+)}=&LA\hat{x}_{k}^{(+)}+LBu_{k}+Ky_{k+1} \\
L=&P_{\infty}^{(+)}E^{T}(P_{\infty}^{(-)})^{-1},~~K=P_{\infty}^{(+)}H^{T}R^{-1}\nonumber  \\
P_{\infty}^{(-)}=& AP_{\infty}^{(+)}A^{T}+Q \nonumber
\end{align}
is stable.
\end{thm}

\begin{pf}
See Theorem 4.3 of \cite{Nikoukhah1992} for proof of the general case when $R\ge 0$ and when both $w_{k}$ and $v_{k}$ are possibly correlated.  Our less general case follows by setting $R>0$ and $S=0$ in the same proof.  Note that \eqref{riccati1} can be expressed as:
\begin{align*}
P_{\infty}^{(+)}=&(LA)P_{\infty}^{(+)}(LA)^{T}+\left(\begin{array}{cc}L&K\end{array}\right)\left(\begin{array}{cc}Q&0\\0&R\end{array}\right)\left(\begin{array}{c}L^{T}\\K^{T}\end{array}\right)
\end{align*}
which is consistent with equation (4.35) in \cite{Nikoukhah1992}.
\end{pf}

\begin{cor} \label{cor1}
The Kalman filter recursion \eqref{kal} converges to the true value of the state $x_{k}^{*}$ when $w_{k}=v_{k}=0$.
\end{cor}

\begin{pf}
As shown in \cite{Nikoukhah1992}, the error dynamics can be expressed as:
\begin{align}
\tilde{x}_{k+1}=&LA\tilde{x}_{k}+Lw_{k}-Kv_{k} \label{xtilde}
\end{align}
where, $\tilde{x}_{k}=x_{k}-\hat{x}_{k}^{(+)}$.  Since $LA$ is stable from Theorem 4.3 of \cite{Nikoukhah1992}, then $\tilde{x}_{k}\rightarrow 0$ as $k\rightarrow \infty$ when $w_{k}=v_{k}=0$ and therefore $\hat{x}_{k}^{(+)}\rightarrow x_{k}^{*}$
\end{pf}

\begin{lem}\label{equalzero}
\begin{align}
\lim _{T\rightarrow \infty}\bar{Z}_{T-N}^{mh}(x_{T-N}^{*})=0
\end{align}
\end{lem}
\begin{pf}
This follows from Corollary \ref{cor1} since $\hat{x}_{T-N-1}^{(+)}\rightarrow x_{T-N-1}^{*}$ and $P_{T-N-1}^{(-)} \rightarrow P_{\infty}^{(-)}$ as $T\rightarrow \infty$, then:
\begin{align*}
&\lim _{T\rightarrow \infty}\bar{Z}_{T-N}^{mh}(x_{T-N}^{*})=\\
&\lim _{T\rightarrow \infty}\|Ex_{T-N}^{*}-A\hat{x}_{T-N-1}^{(+)}-Bu_{T-N-1}\|^{2}_{P_{T-N-1}^{(-)}}=0
\end{align*}
\end{pf}

\begin{lem}\label{up1}
Let $\widehat{J}_{T}^{mhe}$ given by \eqref{mhe1}, then $\widehat{J}_{T}^{mh}\le \bar{Z}_{T-N}^{mh}(x_{T-N}^{*})$
\end{lem}
\begin{pf}
Since the true state sequence $x_{k}^{*}$ is feasible, then: $\widehat{J}_{T}^{mh}\le J_{T}^{mh}(\{x_{k}^{*}\}_{T-N}^{T})$.  But since $Ex_{k+1}^{*}-Ax_{k}^{*}-Bu_{k}=0$ and $y_{k}-Hx_{k}^{*}=0$ (assuming $w_{k}=v_{k}=0$) then $J_{T}^{mh}(\{x_{k}^{*}\}_{T-N}^{T})=\bar{Z}_{T-N}^{mh}(x_{T-N}^{*})$ and the result follows.
\end{pf}

Reformulating the moving horizon optimal cost using \eqref{reformulate1} it can be shown that:
\begin{align}
\widehat{J}_{T}^{mh}&=\sum_{k=T-N}^{T-1}\|\hat{x}_{k}^{mh}-\hat{x}_{k}^{sm}(x_{k+1})\|_{\Gamma_{k}^{sm}}^{2}+\|\hat{x}_{T}^{mh}-\hat{x}_{T}^{(+)}\|_{P_{T}^{(+)}}^{2} \label{mherefor}
\end{align}
From lemma \eqref{up1} and corollary \eqref{equalzero} we can write:
\begin{align*}
\lim_{T\rightarrow \infty}\widehat{J}_{T}^{mh}& \le \lim_{T\rightarrow \infty}\bar{Z}_{T-N}^{mh}(\hat{x}_{T-N}^{*})=0 
\end{align*}
Hence from \eqref{mherefor} and the above inequality, we can conclude that $\|\hat{x}_{T}^{mh}-\hat{x}_{T}^{(+)}\|_{P_{T}^{(+)}}^{2} \rightarrow 0~~\mbox{ as } T\rightarrow \infty$.  Furthermore, since, $P_{k}^{(+)}\rightarrow P_{\infty}^{(+)}>0$ then $\hat{x}_{T}^{mh}\rightarrow \hat{x}_{T}^{(+)}$ and consequently $\hat{x}_{T}^{mh}\rightarrow x_{T}^{*}$ as $T\rightarrow \infty$.

Applying the observer asymptotic stability definition \ref{stabdef}: we assume that the initial term $\|Ex_{1}-A\hat{x}_{0}^{(+)}-Bu_{T}\|\le \delta$, where $\hat{x}_{0}^{(+)}\in \mathbb{X}$, then $\widehat{J}_{T}^{mh}\le \delta^{2}$.  Consequently, from the convergence result above we can find an $\epsilon$ such that $\|\hat{x}_{T}^{mh}-x_{T}\| \le \epsilon$ for all $T\ge \tilde{T}$.  Furthermore, since $\hat{x}_{T}^{mh}\rightarrow x_{T}^{*}$ as $T\rightarrow \infty$ then the MHE is an asymptotically stable observer.

\section{\textbf{Proof of Theorem \eqref{condnum}}}
Assuming the system is detectable and stabilizable, then the sequence $\Gamma_{k}^{sm}$ will converge to the finite positive definite solution corresponding to the solution of the algebraic Riccati equation \eqref{riccati1}:
\begin{align} \label{riccati}
\Gamma_{\infty}^{sm}=&((P_{\infty}^{(+)})^{-1}+A^{T}Q^{-1}A)^{-1}
\end{align}
In other words, as $b_{s},a_{(s+1)}-1 \rightarrow \infty$, where $a_{(s+1)}-1\ge b_{s}$, and: $\|(\Gamma_{b_{s}+1}^{sm})^{-1}M_{c_{s}}\|_{2}\rightarrow
\|(\Gamma_{\infty}^{sm})^{-1}(\Gamma_{\infty}^{sm}A^{T}Q^{-1}E)^{c_{s}}\|_{2}$.
Hence, the stability of the matrix $\Gamma_{\infty}^{sm}A^{T}Q^{-1}E$ is sufficient for the convergence of \eqref{converge} to zero.  Let $M:=\Gamma_{\infty}^{sm}A^{T}Q^{-1}E$, we can rewrite $M$ as follows:
\begin{align*}
M=&((P_{\infty}^{(+)})^{-1}+A^{T}Q^{-1}A)^{-1}A^{T}Q^{-1}E\\
=&((P_{\infty}^{(+)}-P_{\infty}^{(+)}A^{T}(Q+AP_{\infty}^{(+)}A^{T})^{-1}AP_{\infty}^{(+)})A^{T}Q^{-1}E\\
=&P_{\infty}^{(+)}A^{T}Q^{-1}E-P_{\infty}^{(+)}A^{T}(P_{\infty}^{(-)})^{-1}AP_{\infty}^{(+)}A^{T}Q^{-1}E\\
=&P_{\infty}^{(+)}A^{T}(I-(P_{\infty}^{(-)})^{-1}AP_{\infty}^{(+)}A^{T})Q^{-1}E\\
=&P_{\infty}^{(+)}A^{T}((P_{\infty}^{(-)})^{-1}(P_{\infty}^{(-)}) -(P_{\infty}^{(-)})^{-1}AP_{\infty}^{(+)}A^{T})Q^{-1}E\\
=&P_{\infty}^{(+)}A^{T}(P_{\infty}^{(-)})^{-1}E
\end{align*}
We may now rewrite $P_{\infty}^{(+)}$ in terms of $M$ as follows:
\begin{align}
(P_{\infty}^{(+)})^{-1}=&E^{T}(P_{\infty}^{(-)})^{-1}E+H^{T}R^{-1}H\nonumber \\
=&E^{T}(P_{\infty}^{(-)})^{-1}(P_{\infty}^{(-)})(P_{\infty}^{(-)})^{-1}E+H^{T}R^{-1}H\nonumber \\
=&E^{T}(P_{\infty}^{(-)})^{-1}AP_{\infty}^{(+)}A^{T}(P_{\infty}^{(-)})^{-1}E \nonumber \\
&+E^{T}(P_{\infty}^{(-)})^{-1}Q(P_{\infty}^{(-)})^{-1}E+H^{T}R^{-1}H \nonumber \\
=&M^{T}(P_{\infty}^{(+)})^{-1}M+\bar{Q} \label{PD} \\
\mbox{where, } \bar{Q}=&E^{T}(P_{\infty}^{(-)})^{-1}Q(P_{\infty}^{(-)})^{-1}E+H^{T}R^{-1}H \nonumber 
\end{align}
Since $(P_{\infty}^{(-)})^{-1}>0$ and $R^{-1}>0$ then $\bar{Q}>0$.  Furthermore, since $(P_{\infty}^{(+)})^{-1}>0$ then by Lyaponov, \eqref{PD} implies that $M$ is stable.  Hence, the matrix norm sequence \eqref{coupling} converges to zero with increasing value of the time gap $c_{s}$ and since:
\begin{align*}
&\|(\Gamma_{\infty}^{sm})^{-1}(\Gamma_{\infty}^{sm}A^{T}Q^{-1}E)^{a_{(s+1)}-1}\|_{2} 
\\
&~~~~~~~~\le \|(\Gamma_{\infty}^{sm})^{-1}\|_{2}.\|(\Gamma_{\infty}^{sm}A^{T}Q^{-1}E)^{a_{(s+1)}-1}\|_{2}
\end{align*}
and $\Gamma_{\infty}^{sm}< \infty$ the result follows.
\bibliographystyle{plain}        
\bibliography{references}           

\end{document}

%% file: fig1.tex
\tikzfading[name=fade left, left color=transparent!100, right color=transparent!0]

\newcommand*\mystrut[1]{\vrule width0pt height0pt depth#1\relax}

\begin{figure}[t]
\begin{center}
\resizebox{3.0in}{!}{
\begin{tikzpicture}
\draw[->] (0,0) -- ++(7,0) node[below right] {$k$};
\draw[->] (0,0) -- ++(0,1.5) node[above left] {$\hat{x}_{k|T}$};
\draw (0,1.2) node[left] {$x_{max}$} -- ++(0.1,0);
\draw (0,0.25) node[left] {$x_{min}$} -- ++(0.1,0);

\draw (0.3,-.1) node[below] {\small $T-N$};
\draw (0.5,-.1) -- ++(0,.2);
\draw (1.0,-.1) node[below] {\small $T$} -- ++(0,.2);
\draw[domain=0:1.0,smooth,dotted,variable=\t] plot (\t,{0.5*sin(80*\t+20)+0.75});
\draw[fill=blue!20,draw=none,semitransparent] (0.5,.25) rectangle (1.0,1.25);
\draw (1.6,1.2) node[above] { \sffamily \small MHE Sliding Window};
\draw (1.3,1.3) -- ++(-0.3,-0.3);

\draw[->,>=stealth] (1.0,.75) -- ++(.5,0);

\end{tikzpicture}
}
\resizebox{3.0in}{!}{
\begin{tikzpicture}
\draw[->] (0,0) -- ++(7,0) node[below right] {$k$};
\draw[->] (0,0) -- ++(0,1.5) node[above left] {$\hat{x}_{k|T}$};
\draw (0,1.2) node[left] {$x_{max}$} -- ++(0.1,0);
\draw (0,0.25) node[left] {$x_{min}$} -- ++(0.1,0);
\draw (0.75,-.1) node[below] {\small $a_{1}$} -- ++(0,.2);
\draw (1.07,-0.03) node[below] {\small $b_{1}$};
\draw (1.0,-.1) -- ++(0,.2);
\draw (2.3,-.1) node[below] {\small $T-N$};
\draw (2.5,-.1) -- ++(0,.2);
\draw (3.0,-.1) node[below] {\small $T$} -- ++(0,.2);
\draw[domain=0:3.0,smooth,dotted,variable=\t] plot (\t,{0.5*sin(80*\t+20)+0.75});
\draw[fill=blue!20,draw=none,semitransparent] (2.5,.25) rectangle (3.0,1.25);
\draw[fill=purple!20,draw=none,semitransparent] (0.75,.25) rectangle (1.,1.25);
\draw (2.2,1.2) node[above] { \sffamily \small Temporary Fixed Window 1};
\draw (1.3,1.3) -- ++(-0.3,-0.3);
\draw[->,>=stealth] (3.0,.75) -- ++(.5,0);
\end{tikzpicture}
}

\resizebox{3.0in}{!}{
\begin{tikzpicture}
\draw[->] (0,0) -- ++(7,0) node[below right] {$k$};
\draw[->] (0,0) -- ++(0,1.5) node[above left] {$\hat{x}_{k|T}$};
\draw (0,1.2) node[left] {$x_{max}$} -- ++(0.1,0);
\draw (0,0.25) node[left] {$x_{min}$} -- ++(0.1,0);
\draw (0.75,-.1) node[below] {\small $a_{1}$} -- ++(0,.2);
\draw (1.07,-0.03) node[below] {\small $b_{1}$};
\draw (1.0,-.1) -- ++(0,.2);
\draw (4.3,-.1) node[below] {\small $T-N$};
\draw (4.5,-.1) -- ++(0,.2);
\draw (5.0,-.1) node[below] {\small $T$} -- ++(0,.2);
\draw (3.0,-.1) node[below] {\small $a_{2}$} -- ++(0,.2);
\draw (3.32,-.03) node[below] {\small $b_{2}$};
\draw (3.25,-.1) -- ++(0,.2);

\draw[domain=0:5.0,smooth,dotted,variable=\t] plot (\t,{0.5*sin(80*\t+20)+0.75});
\draw[fill=blue!20,draw=none,semitransparent] (4.5,.25) rectangle (5.0,1.25);
\path (0.75,0.25) rectangle (1.0,1.25);
\fill [purple!20,path fading=fade left] (0.75,0.25) rectangle (1,1.25);

\draw (1.6,1.2) node[above] { \sffamily \small Vanishing Window 1};
\draw (1.3,1.3) -- ++(-0.3,-0.3);
\draw[->,>=stealth] (5.0,.75) -- ++(.5,0);

\draw[fill=purple!20,draw=none,semitransparent] (3,.25) rectangle (3.25,1.25);
\draw (3.55,1.3)-- ++(-0.3,-0.3);
\draw (5.2,1.2) node[above] { \sffamily \small Temporary Fixed Window 2};

\end{tikzpicture}
}

\resizebox{3.0in}{!}{
\begin{tikzpicture}
\draw[->] (0,0) -- ++(7,0) node[below right] {$k$};
\draw[->] (0,0) -- ++(0,1.5) node[above left] {$\hat{x}_{k|T}$};
\draw (0,1.2) node[left] {$x_{max}$} -- ++(0.1,0);
\draw (0,0.25) node[left] {$x_{min}$} -- ++(0.1,0);

\draw (4.7,-.1) node[below] {\small $T-N$};
\draw (5.0,-.1) -- ++(0,.2);
\draw (5.5,-.1) node[below] {\small $T$} -- ++(0,.2);
\draw (3.0,-.1) node[below] {\small $a_{2}$} -- ++(0,.2);
\draw (3.32,-.03) node[below] {\small $b_{2}$};
\draw (3.25,-.1) -- ++(0,.2);

\draw[domain=0:5.5,smooth,dotted,variable=\t] plot (\t,{0.5*sin(80*\t+20)+0.75});
\draw[fill=blue!20,draw=none,semitransparent] (5.0,.25) rectangle (5.5,1.25);
\path (3,0.25) rectangle (3.25,1.25);
\fill [purple!20,path fading=fade left] (3.0,0.25) rectangle (3.25,1.25);
\draw (3.55,1.3)-- ++(-0.3,-0.3);
\draw (4.5,1.2) node[above] { \sffamily \small Vanishing Window 2};
\draw[->,>=stealth] (5.5,.75) -- ++(.5,0);

\end{tikzpicture}
}


\caption{Multiple Window Formulation}\label{fig1}
\end{center}
\end{figure}


%% file: fig2.tex
\begin{figure}[t]
\begin{center}
\resizebox{3.5in}{!}{
\begin{tikzpicture}
\draw[->] (0,0) -- ++(7,0) node[below right] {$k$};
\draw[->] (0,0) -- ++(0,2) node[above left] {$\hat{x}_{k|T}$};
\draw (4,-.1) node[below] {$T-N$} -- ++(0,.2);
\draw (6.5,-.1) node[below] {$T$} -- ++(0,.2);
\draw[domain=0:6.5,smooth,dotted,variable=\t] plot (\t,{0.5*cos(80*\t- 70)+0.75});
\draw[fill=blue!20,draw=none,semitransparent] (4,.25) rectangle (6.5,1.25);
\draw (5.25,1.25) node[above] {\scriptsize$\overbrace{\rule{2.5cm}{0pt}}^{\displaystyle\sum_{k=T-N}^{T-1}\|w_{k}\|_{Q}^{2}+\|v_{k}\|_{R}^{2}}$};
\draw[->,>=stealth] (6.5,.75) -- ++(.5,0);
\newcommand*\mystrut[1]{\vrule width0pt height0pt depth#1\relax}

\draw (2,1.25) node[above] {$\overbrace{\rule{4cm}{0pt}}^{\mystrut{2.5ex} Z_{T-N}(x_{T-N})}$};
\end{tikzpicture}
}
\caption{Information Coverage in MHE}\label{fig2}
\end{center}
\end{figure}

%% file: fig3.tex
\begin{figure}[t]
\begin{center}
\resizebox{3.5in}{!}{
\begin{tikzpicture}
\draw[->] (0,0) -- ++(7,0) node[below right] {$k$};
\draw[->] (0,0) -- ++(0,2) node[above left] {$\hat{x}_{k|T}$};

\draw (2.4,-.1) node[below] {$a_{s}$} -- ++(0,.2);
\draw (2.9,-.03) node[below] {$b_{s}$};
\draw (2.9,-.1) -- ++(0,.2);

\draw (5.5,-.1) node[below] {\small $T-N$};
\draw (5.5,-.1) -- ++(0,.2);
\draw (6.5,-.1) node[below] {\small $T$} -- ++(0,.2);
\draw[domain=0:6.5,smooth,dotted,variable=\t] plot (\t,{0.5*cos(80*\t- 30)+0.75});
\draw[fill=blue!20,draw=none,semitransparent] (5.5,.25) rectangle (6.5,1.25);
\draw[fill=purple!20,draw=none,semitransparent] (2.4,.25) rectangle (2.9,1.25);

\newcommand*\mystrut[1]{\vrule width0pt height0pt depth#1\relax}

\draw[->,>=stealth] (6.5,.75) -- ++(.5,0);
\draw (1.1,1.25) node[above] {$\overbrace{\rule{2.1cm}{0pt}}^{\mystrut{1.5ex} IC}$};
\draw (2.65,1.25) node[above] {$\overbrace{\rule{0.6cm}{0pt}}^{\mystrut{1.5ex} FC_{S}}$};
\draw (4.2,1.25) node[above] {$\overbrace{\rule{2.1cm}{0pt}}^{\mystrut{1.5ex} UC_{S}}$};
\draw (6,1.25) node[above] {$\overbrace{\rule{1.1cm}{0pt}}^{\mystrut{1.5ex} SC}$};

\end{tikzpicture}
}
\caption{Multiple Window Formulation}\label{fig3}
\end{center}
\end{figure}

%% file: fig4.tex
\tikzstyle{fancytitle} =[text=black]    
\begin{figure*}[t]
\begin{center}
\resizebox{6.5in}{!}{{\scalefont{1.3}
\begin{tikzpicture}
[block/.style={draw,rectangle,line width=3pt,inner sep=4pt,outer sep=1pt,minimum width=1.5cm,minimum height=1.5cm},
init/.style={draw,rectangle,rounded corners=10pt,line width=3pt,inner sep=2pt,outer sep=1pt,minimum width=1.5cm,minimum height=0.75cm},
split/.style={draw,diamond,aspect=1.5,line width=3pt,inner sep=2pt,outer sep=1pt,minimum width=1.5cm,minimum height=0.75cm}]

\draw (0,1.5) node[init,draw=violet] (Init0) {\sffamily \textbf{Start}};
\draw (0,-0.8) node[block,draw=violet] (Init1)  {\begin{minipage}{5.5in}\sffamily\setlist[enumerate,1]{leftmargin=0.6cm}
\begin{enumerate}[1)]
\item Solve \eqref{full0} up to time $T=N$ given $\bar{x}_{0},P_{0}$
\item Find $\hat{x}_{T-N}^{(+)},P_{T-N}^{(-)}$ using \eqref{recur} with $\hat{x}_{0}^{(+)}=\bar{x}_{0},~P_{0}^{(+)}=P_{0}$
\item Set New window flag ``on'', $FC=UC=0$, $s_{min}=1$, $S=0$, $a_{1}=b_{1}=0$
\item Extract $\hat{x}_{T|T}^{mw}$ and $\hat{x}_{T-N|T}^{mw}$ from $\hat{\zeta}$, Update $\mathcal{A}(T-N)$ \eqref{actset}
\end{enumerate}\end{minipage}};
\node[fancytitle] at (-5.8,1) {\sffamily\textbf{Initialization}};

\draw (-2,-3.2) node[circle,draw] (goto1) {\textbf 1};

\draw (0,-4.5) node[split,draw=gray,inner sep=-1pt,aspect=3] (ExtendWindowCheck) {\begin{minipage}{1in}\centering\sffamily $\mathcal{A}(T-N)= \{\emptyset \}$?\end{minipage}};

\draw (1,-9) node[split,draw=gray,aspect=2] (OldWindowCheck) {\begin{minipage}{1.6in}\textbf\normalsize\centering\sffamily Is $S\ge s_{min}$ and $T>b_{s_{min}} + N_{FC}+1$?\end{minipage}} ;

\draw (9,-14) node[block,draw=green] (Run)  {\begin{minipage}{5in}\sffamily\setlist[enumerate,1]{leftmargin=0.6cm}
\begin{enumerate}[1)]\setlength{\baselineskip}{20pt}
\item Set $T=T+1$, 
\item Reconstruct $FC_{i},UC_{i}^{(-)}$ for $s=s_{min},\cdots,S-1$ from memory
\item Calculate and store $\hat{x}_{T-N}^{(+)},~P_{T-N}^{(-)}$ using  $\hat{x}_{T-N-1}^{(+)},P_{T-N-1}^{(-)}$ \eqref{recur}
\item Construct $\bar{Z}_{T-N}^{mh}(x_{T-N})$ \eqref{arrivalsel}
and $SC(\{x_{k}\}_{T-N}^{T})$ \eqref{smooth2}
\item Solve \eqref{mwmhe}; Extract $\hat{x}_{T|T}^{mw}$ and $\hat{x}_{T-N|T}^{mw}$; Update $\mathcal{A}(T-N)$ \eqref{actset}
\end{enumerate}\end{minipage}};
\node[fancytitle] at (9,-11.7) {\sffamily\textbf{Solve All Windows}};

\draw (10,-4.5) node[block,draw=cyan] (NotActive) {\begin{minipage}{4.65in}\sffamily\setlist[enumerate,1]{leftmargin=0.6cm}
\begin{enumerate}[1)]
\item Set New Window flag ``on'', $a_{(S+1)} = T-N, c_{S}=a_{(S+1)}-b_{S}$
\item If $S\ge s_{min}$ 
\begin{enumerate}[leftmargin=0.2cm]
\item Calculate $\&$ store $M_{c_{S}}^{S}$ and $r_{c_{S}-1}^{S}$ \eqref{Mi}, \eqref{ri}
\item Construct $\hat{x}_{a_{(S+1)}}^{sm}(x_{T-N})$ and $\hat{x}_{b_{S}+1}^{sm}(x_{a_{(S+1)}}-1)$  \eqref{smooth2},\eqref{xtau2}
\item Construct $UC_{S}^{(-)}$ \eqref{uncon} 
\end{enumerate}
\end{enumerate}\end{minipage}};
\node[fancytitle] at (10,-2.5) {\sffamily\textbf{Detach First Window and Grow $UC_{S}^{(-)}$}};

\draw (10,-9) node[block,draw=red] (ElimWindow)  {\begin{minipage}{3.8in}\sffamily\setlist[enumerate,1]{leftmargin=0.6cm}
\begin{enumerate}[1)]
\item Elliminate $FC_{s_{min}},UC_{s_{min}}$
\item Clear memory $\{\hat{x}_{k}^{(+)},P_{k}^{(+)},\Gamma_{k}^{sm},~k\le b_{smin}\}$
\item Clear memory $\{M_{i}^{s_{min}},r_{i}^{s_{min}}\}$, $i=1,\cdots,c_{s_{min}}$
\item $s_{min}\leftarrow s_{min} + 1$
\end{enumerate}\end{minipage}};
\node[fancytitle] at (10,-7.3) {\sffamily\textbf{Elliminate Last Window}};

\draw (16.5,-14) node[circle,draw] (goto2) {\textbf 1};

\draw (-8,-4.5) node[split,draw=gray,inner sep=0pt,aspect=3] (FlagCheck) {\begin{minipage}{1in}\centering\sffamily Is new Window flag On?\end{minipage}};

\draw (-8,-9) node[block,draw=blue] (NewWindow)  {\begin{minipage}{3.2in}\sffamily\setlist[enumerate,1]{leftmargin=0.6cm}
\begin{enumerate}[1)]
\item $S \leftarrow S+1$, $a_{S}=T-N$, $b_{S}=T-N-1$ 
\item Set New Window flag ``off''
\item $UC_{S}^{(-)} = 0,~M_{1}^{S}=I$, and
\end{enumerate}
$r_{1}^{S}=\hat{x}_{b_{S}+1}^{(+)}-\Gamma_{b_{S}+1}^{sm}A^{T}Q^{-1}(A\hat{x}_{b_{S}+1}^{(+)}+Bu_{b_{S}+1})$
\end{minipage}};
\node[fancytitle] at (-10.2,-7.2) {\sffamily\textbf{Form New Window}};

\draw (-8,-14.2) node[block,draw=blue] (FCupdate)  {\begin{minipage}{3.2in}\sffamily\setlist[enumerate,1]{leftmargin=0.6cm}
\begin{enumerate}[1)]
\item $b_{S} \leftarrow b_{S}+1$
\item Calculate $\&$ store $\Gamma_{b_{S}}^{sm}$ using $P_{b_{S}}^{(+)}$ \eqref{smooth2}
\item Construct $\hat{x}_{b_{S}}^{sm}(x_{b_{S}+1})$ using $\hat{x}_{b_{S}}^{(+)}$ \eqref{smooth2}
\item Construct $FC_{S}$ using $\{\hat{x}_{k}^{sm}(x_{k+1}),\Gamma_{k}^{sm}\}$ for $k=a_{S},\cdots,b_{S}$ \eqref{FC}
\end{enumerate}\end{minipage}};
\node[fancytitle] at (-10.8,-12.2) {\sffamily\textbf{Grow Window}};

\draw[->,ultra thick] (Init0.-90) -- (Init1.90);
\draw[->,ultra thick] (Init1.-90) -- (ExtendWindowCheck.90);
\draw[->,ultra thick]  (goto1.0) -- (ExtendWindowCheck.90 |- goto1.0);
\draw[->,ultra thick] (ExtendWindowCheck.0) -- node[pos=.1,above] {\sffamily yes} (NotActive.180);
\draw[->,ultra thick] (ExtendWindowCheck.180) -- node[pos=.1,above] {\sffamily no} (FlagCheck.0);
\draw[<-,ultra thick] (OldWindowCheck.90) -- ++(0,0.5);
\draw[<-,ultra thick] (OldWindowCheck.90) ++(0,0.5) -| (NotActive.-90);
\draw[<-,ultra thick] (OldWindowCheck.90) ++(0,0.5) -- ++(-3.5,0) |-  (FCupdate.0);
\draw[->,ultra thick] (OldWindowCheck.0) -- node[pos=.1,above] {\sffamily yes} (ElimWindow.180);
\draw[->,ultra thick] (OldWindowCheck.-90) -- node[pos=.25,left] {\sffamily no} ++ (0,-3.35) -- (Run.180);
\draw[<-,ultra thick] (OldWindowCheck) ++(0,-1.9) -| (ElimWindow.-90);
\draw[->,ultra thick] (Run.0) -- (goto2.180);

\draw[->,ultra thick] (FlagCheck.-90) -- node[pos=.3,right] {\sffamily yes} (NewWindow.90);
\draw[->,ultra thick] (NewWindow.-90) -- (FCupdate.90);
\draw[<-,ultra thick] (FCupdate.90) ++(0,1.0) -- ++(-4.7,0) |- node[pos=.9,above] {\sffamily no} (FlagCheck.-180);
\end{tikzpicture}}}
\caption{{Multiple Window Moving Horizon Estimation Aglorithm}}\label{fig4}
\end{center}
\end{figure*}